\documentclass[aps,pre,twocolumn,floats,showpacs,superscriptaddress]{revtex4}
\usepackage{epsfig}

\newcommand{\avk}{\left< k \right>}
\newcommand{\fluck}{\left< k^2 \right>}
\newcommand{\kcmL}{\left( \frac{k_c}{m} \right)}
\newcommand{\kcm}{(k_c / m)}
\begin{document}

\title{Epidemic dynamics in finite size scale-free networks}

\author{Romualdo Pastor-Satorras}
\affiliation{Departament  de F{\'\i}sica i Enginyeria Nuclear,
  Universitat Polit{\`e}cnica de Catalunya, Campus Nord B4,
  08034 Barcelona, Spain}
\author{Alessandro Vespignani}
\affiliation{The Abdus Salam International Centre for Theoretical Physics
  (ICTP), P.O. Box 586, 34100 Trieste, Italy}

\date{\today}

\begin{abstract}
  Many real networks present a bounded scale-free behavior with a
  connectivity cut-off due to physical constraints or a finite network
  size. We study epidemic dynamics in bounded scale-free networks with
  soft and hard connectivity cut-offs. The finite size effects
  introduced by the cut-off induce an epidemic threshold that
  approaches zero at increasing sizes. The induced epidemic threshold
  is very small even at a relatively small cut-off, showing that the
  neglection of connectivity fluctuations in bounded scale-free
  networks leads to a strong over-estimation of the epidemic
  threshold.  We provide the expression for the infection prevalence
  and discuss its finite size corrections. The present work shows that
  the highly heterogeneous nature of scale-free networks does not
  allow the use of homogeneous approximations even for systems of
  a relatively small number of nodes.
\end{abstract}

\maketitle

In the last years it has been recognized that a large number of
physical, biological, and social networks exhibits complex topological
properties~\cite{amaral,strog01}.  In particular many real world
networks show the small-world phenomenon, related to a very small
average path length between nodes~\cite{strog01,watts98}.  More
strikingly, in some cases this property is associated to a scale-free
connectivity distribution, $P(k)\sim k^{-2-\gamma}$, with
$0<\gamma\leq 1$, where $k$ is the number of links connected to a node
\cite{barab99}.  This scale-free nature is associated to a large
heterogeneity in the connectivity properties of the system. Since the
second moment of the connectivity distribution $\fluck$ is diverging
when increasing the network size, the connectivity fluctuations in
scale-free (SF) networks do not have an intrinsic bound and diverge in
the infinite system size limit.  Scale-free properties have been
observed in many real systems such as the
Internet~\cite{falou99,calda00,alexei} and the
World-Wide-Web~\cite{barab99,www99}, food-webs, protein, and neural
networks~\cite{barabasi01}. A very important example of scale-free
networks is also found in the web of human sexual
contacts~\cite{amaral01}.  This is a particularly relevant case since
the unambiguous definition of contacts (links) is often missing in the
analysis of social networks.

Since the Internet and the web of human sexual contacts appear to be
scale-free, the study of epidemics and disease dynamics on SF networks
is a relevant theoretical issue in the spreading of computer viruses
and sexually transmittable diseases. In heterogeneous networks, it is
well-known that the epidemic threshold decreases with the standard
deviation of the connectivity distribution~\cite{het84,anderson92},
and this feature is amplified in SF networks, which have diverging
connectivity fluctuations in the limit of infinite network size.
Indeed, it was first noted in Ref.~\cite{pv01avir} that, in
infinite SF networks, epidemic processes do not possess an epidemic
threshold below which diseases cannot produce a major epidemic
outbreak or the inset of an endemic state.
The absence of an intrinsic epidemic threshold has been found in both
the susceptible-infected-susceptible (SIS) model~\cite{pv01avir}
and the susceptible-infected-removed (SIR)
model~\cite{lloydsir,moreno} in infinite SF networks.  The
immunization policies are as well very much affected by the SF nature
of the connectivity distribution~\cite{psvpro,aidsbar}.

As customarily encountered in nonequilibrium statistical
systems~\cite{marro99}, it has also been pointed out that in finite
systems an epidemic threshold is induced by finite size
effects~\cite{lloydsir}.  Real systems are actually made up by a
finite number of individuals which is far from the thermodynamic
limit.  This finite population introduces a maximum connectivity
$k_c$, depending on $N$, which has the effect of restoring a bound in
the connectivity fluctuations, inducing in this way an effective
nonzero threshold. More generally, we can consider a class of bounded
scale-free (BSF) networks, in which the connectivity distribution has
the form $P(k) \sim k^{-2-\gamma} f(k/k_c)$, where the function $f(x)$
decreases very rapidly for $x>1$ \cite{dorogorev}. The cut-off $k_c$
can be due to the finite size of the network or to the presence of
constraints limiting the addition of new links in an otherwise
infinite network ~\cite{amaral}.

In this paper we present an analytical study of the SIS model in BSF
networks with a generic connectivity exponent $\gamma$ ($0<\gamma\leq1$),
focusing on the effects introduced by a finite cut-off $k_c$. We
analyze the case of a hard cut-off, $f(x) = \theta(1-x)$, where $\theta(x)$ is
the Heaviside step function, as it happens in growing networks with a
finite number of elements.  We consider as well a soft exponential
cut-off, $f(x) = \exp(-x)$, as often found in systems where physical
constraints are at play.  We derive the behavior of the epidemic
threshold as a function of $k_c$ and the network size $N$, and find
that even for relatively small networks the induced epidemic threshold
is much smaller than the epidemic threshold found in homogeneous
systems.  This confirms that the SF nature cannot be neglected in the
practical estimates of epidemic and immunization thresholds in real
networks. We also provide the explicit analytic form for the epidemic
prevalence (density of infected individuals) in BSF networks. The
results presented here can be readily extended to the SIR case.

\begin{figure}[t]
  \centerline{\epsfig{file=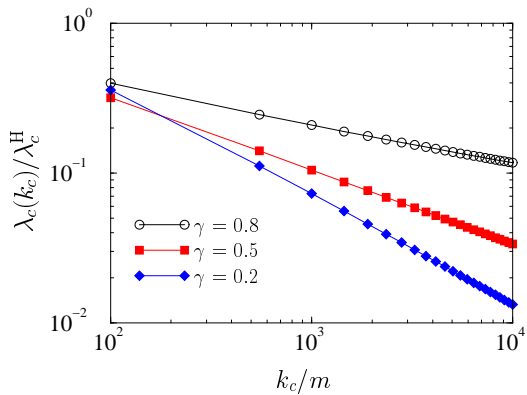, width=2.7in}} 
  \caption{Ratio between the effective epidemic threshold 
    in BSF networks with a soft exponential cut-off $k_c$  and the
    intrinsic epidemic threshold of homogeneous networks with the same
    average connectivity, for different values of $\gamma$.}
\label{figkc}
\end{figure}

In order to estimate the effect of $k_c$ in epidemics on BSF networks
we will investigate the standard SIS model~\cite{epidemics}.  This
model relies on a coarse-grained description of individuals in the
population.  Namely, each node of the graph represents an individual
and each link is a connection along which the infection can spread.
Each susceptible (healthy) node is infected with rate $\nu$ if it is
connected to one or more infected nodes.  Infected nodes are cured and
become again susceptible with rate $\delta$, defining an effective
spreading rate $\lambda=\nu/\delta$ (without lack of generality, we
set $\delta=1$).  The SIS model does not take into account the
possibility of individuals removal due to death or acquired
immunization \cite{epidemics}, and individuals run stochastically
through the cycle susceptible $\to$ infected $\to$ susceptible.  This
model is generally used to study infections leading to endemic states
with a stationary average density of infected individuals.  In order
to take into account the heterogeneity of SF networks, we have to
relax the homogeneity assumption used in regular networks, and
consider the relative density $\rho_k(t)$ of infected nodes with given
connectivity $k$; i.e., the probability that a node with $k$ links is
infected~\cite{pv01avir}.  The dynamical mean-field equations can
thus be written as
\begin{equation}
  \frac{ d \rho_k(t)}{d t} = -\rho_k(t) +\lambda k \left[
  1-\rho_k(t) \right] \Theta(\rho(t)).
\label{mfk}
\end{equation}
The first term in Eq.~(\ref{mfk}) considers infected nodes becoming
healthy with unit rate. The second term represents the average density
of newly generated infected nodes that is proportional to the
infection spreading rate $\lambda$ and the probability that a node
with $k$ links is healthy $[1-\rho_k(t)]$ and gets the infection via a
connected node. The rate of this last event is given by the
probability $\Theta(\rho(t))$ that any given link points to an
infected node, which has the expression~\cite{pv01avir}
\begin{equation}
  \Theta(\rho(t))= \avk^{-1}\sum_k kP(k)\rho_k(t).
  \label{first}
\end{equation}
By solving Eqs.~(\ref{mfk}) and~(\ref{first}) in the stationary state
[$d \rho_k(t)/ dt =0$] we obtain the self-consistency
equation~\cite{pv01avir}
\begin{equation}
  \Theta = \avk^{-1}  \sum_k k P(k)  \frac{\lambda k \Theta}{1 +
    \lambda k \Theta},
\label{cons}
\end{equation}
where $\Theta$ is now a function of $\lambda$ alone.  The
self-consistency Eq.~(\ref{cons}) allows a solution with $\Theta\neq
0$ and $\rho_k\neq 0$ only if the condition $\lambda \fluck/\avk\geq1$
is fulfilled~\cite{psvpro}, defining the epidemic threshold
\begin{equation}
\lambda_c = \frac{\avk}{\fluck}.
\label{thr}
\end{equation}
In other words, if the value of $\lambda$ is above the threshold,
$\lambda\geq \lambda_c$, the infection spreads and becomes endemic.
Below it, $\lambda <\lambda_c$, the infection dies out exponentially
fast. This result implies that in infinite SF networks with
connectivity exponent $0<\gamma\leq 1$, for which $\fluck\to\infty$,
we have $\lambda_c=0$.  This fact implies in turn that for any
positive value of $\lambda$ the infection can pervade the system with
a finite prevalence, in a sufficiently large
network~\cite{pv01avir}.  While this results is valid for infinite
SF networks, $\fluck$ assumes a finite value in BSF networks, defining
an effective nonzero threshold due to finite size effects as usually
encountered in nonequilibrium phase transitions~\cite{marro99}.  This
epidemic threshold, however, is not an {\em intrinsic} quantity as in
homogeneous systems and it vanishes for a increasing network size or
connectivity cut-off.  In order to calculate the precise effects of a
finite $k_c$ we consider two different cases of connectivity cut-off.
At first instance we consider a {\em soft} exponential cut-off with
{\em characteristic} connectivity $k_c$. This case corresponds to
those real networks in which external factors set up an upper limit to
the connectivity~\cite{amaral}.  The network can have an infinite
number of elements but the power-law connectivity distribution decays
exponentially for large values of $k$.  In order to perform explicit
calculations we use a continuous approximation that substitutes the
connectivity by a real variable $k$ in the range $[m,\infty)$, where
$m$ is the minimum connectivity of the network.  The connectivity
probability distribution in this case is $P(k)=A
k^{-2-\gamma}\exp(-k/k_c)$, where $A$ is a normalization factor.  The
effective nonzero epidemic threshold $\lambda_c(k_c)$ induced by the
exponential cut-off is given by
\begin{equation}
\lambda_c(k_c)=\frac{\int_m^{\infty}k^{-1-\gamma}\exp(-k/k_c) dk}
{\int_m^{\infty}k^{-\gamma}\exp(-k/k_c)dk}, 
\end{equation}
which, after integration,  yields
\begin{equation}
  \lambda_c(k_c)=k_c^{-1} \frac{\Gamma(-\gamma,
    m/k_c)}{\Gamma(1-\gamma,  m/k_c)},
\label{thsoft}
\end{equation}
where $\Gamma(x,y)$ is the incomplete Gamma function
\cite{abramovitz}. For large $k_c$ we can perform a Taylor expansion
and retain only the leading term, obtaining for any $0 <\gamma < 1$
\begin{equation}
\lambda_c(k_c) \simeq \frac{1}{m\gamma\Gamma(1-\gamma)}\kcm^{\gamma-1}.
\end{equation}
The limit $\gamma \to 1$ in Eq.~(\ref{thsoft}) corresponds to a
logarithmic divergence, yielding at leading order
$\lambda_c(k_c)\simeq(m\ln(k_c/m))^{-1}$.  In all cases we have that
the epidemic threshold vanishes when increasing the characteristic
cut-off.  For large $k_c$, the average connectivity is virtually fixed
and given by $\avk= (\gamma +1) m /\gamma$, for any $\gamma > 0$.  It
is interesting thus to compare the intrinsic epidemic threshold
obtained in homogeneous networks with negligible fluctuations and the
nonzero effective threshold of BSF networks.  The intrinsic epidemic
threshold of homogeneous networks with constant node connectivity
$\avk$ is given by $\lambda_c^{\rm
  H}=\avk^{-1}$~\cite{epidemics,pv01avir}.  If we compare BSF and
homogeneous networks with the same average connectivity $\avk=(\gamma
+1) m /\gamma$ we obtain that the ratio between the epidemic
thresholds is given by
\begin{equation}
\frac{\lambda_c(k_c)}{\lambda_c^{\rm H}} \simeq
\frac{(\gamma+1)}{\gamma^2\Gamma(1-\gamma)}\kcm^{\gamma-1}.
\end{equation}
This clearly shows that even in the case of a connectivity cut-off the
effective epidemic threshold in BSF networks is much smaller than the
intrinsic threshold obtained in regular networks.  In Fig.~\ref{figkc}
we plot the ratio obtained by using the full expression for
$\lambda_c(k_c)$, Eq.~(\ref{thsoft}). It is striking to observe that, even
with relatively small cut-offs ($k_c\sim 10^2-10^3$), for $\gamma\approx0.5$ the
effective epidemic threshold of BSF networks is smaller by a factor
close to $1/10$ than the intrinsic threshold obtained on homogeneous
networks.

As a second kind of finite size effect we consider the presence of a
{\em hard} cut-off $k_c$. Since SF networks are often dynamically
growing networks, this case represent a network which has grown up to
a finite number of nodes $N$. The maximum connectivity $k_c$ of any
node is related to the network age, measured as the number of nodes
$N$, by the scaling relation \cite{dorogorev}
\begin{equation}
  k_c \simeq  m N^{1/(1+ \gamma)},
\label{Nscaling}
\end{equation}
where $m$ is the minimum connectivity of the network.

\begin{figure}[t]
  \centerline{\epsfig{file=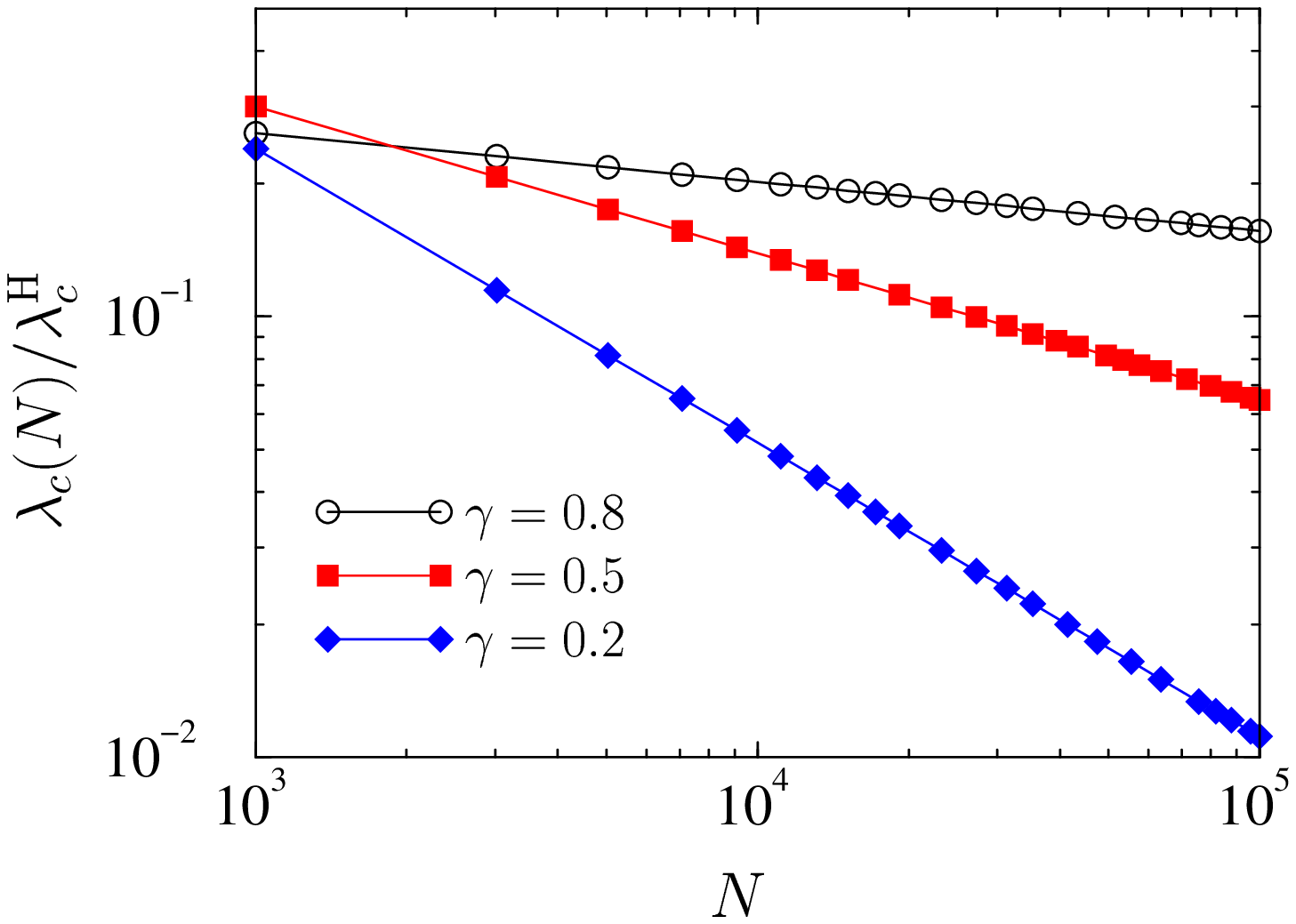, width=2.7in}} 
  \caption{Ratio between the effective epidemic threshold 
    in BSF networks with finite size $N$ and the intrinsic epidemic
    threshold of homogeneous networks with the same average
    connectivity, for differen values of $\gamma$.}
\label{figN}
\end{figure}

In this case the network does not possess any node with connectivity
$k$ larger than $k_c$, and we can think in terms of a hard cut-off.
Using again the continuous $k$ approximation, the normalized
connectivity distribution has now the form
\begin{equation}
P(k) = \frac{(1+\gamma) m^{1+\gamma}}{ 1 -\kcm
^{-1-\gamma}}
  k^{-2-\gamma} \theta(k_c-k),
  \label{eq:dist}
\end{equation}
where $\theta(x)$ is the Heaviside step function.  The finite size
induced epidemic threshold $\lambda_c(k_c)$ is given by the expression
\begin{equation}
\lambda_c(k_c)=\frac{\int_m^{k_{c}}
  k^{-1-\gamma}dk}{\int_m^{k_{c}}k^{-\gamma} dk}.
\end{equation}
Evaluating the above expression we obtain at leading order in $k_c/m$:
\begin{equation}
\lambda_c(k_c) \simeq \frac{1-\gamma}{\gamma m}\kcm^{\gamma-1}.
\label{eq:hardkc}
\end{equation}
In this case the hard cut-off $k_c$ can be expressed as a function of
the network size $N$ by using the scaling relation
Eq.~(\ref{Nscaling}) and we can obtain the effective epidemic
threshold as
\begin{equation}
\lambda_c(N)  \simeq
  \frac{1-\gamma}{\gamma m } N^{(\gamma-1)/(\gamma+1)}.
\label{thres}
\end{equation}
This expression is valid for any $0<\gamma<1$, while for $\gamma=1$ we
obtain at the leading order the logarithmic behavior
$\lambda_c(N)\simeq2(m\ln(N))^{-1}$.  Also in this case we have that
the effective epidemic threshold is approaching zero for increasing
network sizes, and it is worth comparing its magnitude with the
corresponding intrinsic threshold in homogeneous networks with
identical average connectivity. In Fig.~\ref{figN} we report the ratio
$\lambda_c(N)/\lambda_c^{\rm H}$ for different sizes of the SF
network.  It is striking to notice that for $\gamma=0.5$, small
networks with $N\simeq 10^4$ exhibit a finite size induced epidemic
threshold that is close to be one order of magnitude smaller than the
intrinsic epidemic threshold of a homogeneous network.

In order to find the prevalence behavior we have to solve
Eq.~(\ref{cons}) in the continuous approximation,
\begin{equation}
  \Theta = \avk^{-1} \int_m^\infty k P(k) \frac{\lambda \Theta k}{1 +
    \lambda \Theta k}  dk,
  \label{eq:1}
\end{equation}
and use the value of $\Theta$ to compute the density of infected sites
$\rho$ as 
\begin{equation}
  \rho  =  \sum_k \rho_k P(k) \equiv \int_m^\infty  P(k) \frac{\lambda
    \Theta k}{1 + \lambda \Theta k} dk,
    \label{eq:2}
\end{equation}
where $P(k)$ is given by Eq.~(\ref{eq:dist}). In the absence of any
cut-off ($k_c\to\infty$) and in the thermodynamic limit ($N\to\infty$)
the prevalence scales as $\rho\sim\lambda^{1/(1-\gamma)}$ if
$0<\gamma<1$, and as $\rho\sim\exp(-1/m \lambda)$ if
$\gamma=1$~\cite{pv01avir}.  Accordingly with the absence of the
epidemic threshold, the prevalence is null only if the spreading rate
is $\lambda=0$. In the case of a hard cut-off we can integrate
Eq.~(\ref{eq:1}), neglecting terms of order $(k_c/m)^{-\gamma}$ in the
$P(k)$ distribution, to obtain:
\begin{eqnarray*}
  \Theta&\simeq& \gamma m^\gamma  \lambda \Theta 
  \int_m^{k_c} \frac{k^{-\gamma}}{1  + \lambda    \Theta k} dk\\
  &=& F(1,\gamma,1+\gamma,-[\lambda \Theta m]^{-1} )\\ &~&
  -\kcm^{-\gamma} F(1,\gamma,1+\gamma,-[\lambda \Theta k_c]^{-1}), 
\end{eqnarray*}
where $F$ is the Gauss hyper-geometric function \cite{abramovitz}.
For a \textit{fixed} $k_c$ one can expand both hyper-geometric
functions on the right hand side in the previous equation, keeping the
most relevant terms in $\Theta$ and considering afterwards the limit
of large $k_c$. The final solution for $\Theta$ is then given, at
leading order in $\kcm$, by
\begin{equation}
  \Theta \simeq  \frac{1}{m \lambda^2}
  \frac{2-\gamma}{1-\gamma} \kcmL^{-1} \left[ \lambda -
    \frac{1-\gamma}{\gamma m} \kcmL^{\gamma-1}  \right].
  \label{eq:thetaSF}
\end{equation}

By evaluating the integral in Eq.~(\ref{eq:2}) and keeping the 
leading term in $\Theta$ and $k_c$ we finally obtain the infection 
prevalence as 
\begin{displaymath}
  \rho \simeq \frac{(\gamma+1)(2-\gamma)}{\lambda \gamma(1-\gamma) } 
  \kcmL^{-1}  \left[ \lambda -
    \frac{1-\gamma}{\gamma m} \kcmL^{\gamma-1}   \right].
\end{displaymath}
Inserting the scaling relation Eq.~(\ref{Nscaling}) between the
maximum connectivity $k_c$ and the network size $N$ we are led to the
final expression
\begin{equation}
  \rho \sim N^{-1/(\gamma+1)} (\lambda -\lambda_c(N) ).
  \label{eq:prevN}
\end{equation}
That is, the finite size of the network induces a standard mean-field
transition at the induced epidemic threshold $\lambda_c(N)$, given by
Eq.~(\ref{thres}). As can be seen from Eq.~(\ref{eq:prevN}), however,
the prevalence is depressed by a factor $N^{-1/(\gamma+1)}$ from the
corresponding value for a homogeneous network.  The above calculations
can be repeated along similar lines in the case of a {\em soft}
exponential cut-off, obtaining similar results.

It is worth remarking that similar results hold as well for the SIR
model.  Despite this model confers permanent immunity and does not
allow for a stationary state, the epidemic threshold over which an
epidemic outbreak occurs has the same analytic form
$\lambda_c=\avk/\fluck$~\cite{moreno}.  Thus, the present results for the
effect of finite size and the induced epidemic threshold can be
readily exported to the SIR case.  The calculation of the epidemic
prevalence is different due to the different evolution equations, but
recovers the same onset of an induced mean-field transition at the
effective threshold $\lambda_c(N)$. Interestingly, similar results have
been obtained in the analysis of the resilience to damage of finite
size scale-free networks \cite{havlin01,newman00}

In conclusion, we have shown that the SF networks weakness to epidemic
agents is also present in finite size networks. Using the homogeneity
assumption in the case of SF networks will lead to a serious
over-estimate of the epidemic threshold even for relatively small
networks.

This work has been partially supported by the European Network
Contract No. ERBFMRXCT980183.  R.P.-S. acknowledges financial support
from the Ministerio de Ciencia y Tecnolog\'{\i}a (Spain) and from the
Abdus Salam International Centre for Theoretical Physics (ICTP), where
part of this work was done. We thank F. Cecconi for helpful
discussions.


\begin{thebibliography}{22}
\expandafter\ifx\csname natexlab\endcsname\relax\def\natexlab#1{#1}\fi
\expandafter\ifx\csname bibnamefont\endcsname\relax
  \def\bibnamefont#1{#1}\fi
\expandafter\ifx\csname bibfnamefont\endcsname\relax
  \def\bibfnamefont#1{#1}\fi
\expandafter\ifx\csname citenamefont\endcsname\relax
  \def\citenamefont#1{#1}\fi
\expandafter\ifx\csname url\endcsname\relax
  \def\url#1{\texttt{#1}}\fi
\expandafter\ifx\csname urlprefix\endcsname\relax\def\urlprefix{URL }\fi
\providecommand{\bibinfo}[2]{#2}
\providecommand{\eprint}[2][]{\url{#2}}

\bibitem[{\citenamefont{Amaral et~al.}(2000)\citenamefont{Amaral, Scala,
  Barth\'{e}l\'{e}my, and Stanley}}]{amaral}
\bibinfo{author}{\bibfnamefont{L.~A.~N.} \bibnamefont{Amaral}},
  \bibinfo{author}{\bibfnamefont{A.}~\bibnamefont{Scala}},
  \bibinfo{author}{\bibfnamefont{M.}~\bibnamefont{Barth\'{e}l\'{e}my}},
  \bibnamefont{and} \bibinfo{author}{\bibfnamefont{H.~E.}
  \bibnamefont{Stanley}}, \bibinfo{journal}{Proc. Natl. Acad. Sci. USA}
  \textbf{\bibinfo{volume}{97}}, \bibinfo{pages}{11149} (\bibinfo{year}{2000}).

\bibitem[{\citenamefont{Strogatz}(2001)}]{strog01}
\bibinfo{author}{\bibfnamefont{S.~H.} \bibnamefont{Strogatz}},
  \bibinfo{journal}{Nature} \textbf{\bibinfo{volume}{410}},
  \bibinfo{pages}{268} (\bibinfo{year}{2001}).

\bibitem[{\citenamefont{Watts and Strogatz}(1998)}]{watts98}
\bibinfo{author}{\bibfnamefont{D.~J.} \bibnamefont{Watts}} \bibnamefont{and}
  \bibinfo{author}{\bibfnamefont{S.~H.} \bibnamefont{Strogatz}},
  \bibinfo{journal}{Nature} \textbf{\bibinfo{volume}{393}},
  \bibinfo{pages}{440} (\bibinfo{year}{1998}).

\bibitem[{\citenamefont{Barab\'{a}si and Albert}(1999)}]{barab99}
\bibinfo{author}{\bibfnamefont{A.-L.} \bibnamefont{Barab\'{a}si}}
  \bibnamefont{and} \bibinfo{author}{\bibfnamefont{R.}~\bibnamefont{Albert}},
  \bibinfo{journal}{Science} \textbf{\bibinfo{volume}{286}},
  \bibinfo{pages}{509} (\bibinfo{year}{1999}).

\bibitem[{\citenamefont{Faloutsos et~al.}(1999)\citenamefont{Faloutsos,
  Faloutsos, and Faloutsos}}]{falou99}
\bibinfo{author}{\bibfnamefont{M.}~\bibnamefont{Faloutsos}},
  \bibinfo{author}{\bibfnamefont{P.}~\bibnamefont{Faloutsos}},
  \bibnamefont{and}
  \bibinfo{author}{\bibfnamefont{C.}~\bibnamefont{Faloutsos}},
  \bibinfo{journal}{Comput. Commun. Rev.} \textbf{\bibinfo{volume}{29}},
  \bibinfo{pages}{251} (\bibinfo{year}{1999}).

\bibitem[{\citenamefont{Caldarelli et~al.}(2000)\citenamefont{Caldarelli,
  Marchetti, and Pietronero}}]{calda00}
\bibinfo{author}{\bibfnamefont{G.}~\bibnamefont{Caldarelli}},
  \bibinfo{author}{\bibfnamefont{R.}~\bibnamefont{Marchetti}},
  \bibnamefont{and}
  \bibinfo{author}{\bibfnamefont{L.}~\bibnamefont{Pietronero}},
  \bibinfo{journal}{Europhys. Lett.} \textbf{\bibinfo{volume}{52}},
  \bibinfo{pages}{386} (\bibinfo{year}{2000}).

\bibitem[{\citenamefont{Pastor-Satorras
  et~al.}(2001)\citenamefont{Pastor-Satorras, V{\'a}zquez, and
  Vespignani}}]{alexei}
\bibinfo{author}{\bibfnamefont{R.}~\bibnamefont{Pastor-Satorras}},
  \bibinfo{author}{\bibfnamefont{A.}~\bibnamefont{V{\'a}zquez}},
  \bibnamefont{and}
  \bibinfo{author}{\bibfnamefont{A.}~\bibnamefont{Vespignani}},
  \bibinfo{journal}{Phys. Rev. Lett.} \textbf{\bibinfo{volume}{87}},
  \bibinfo{pages}{258701} (\bibinfo{year}{2001}).

\bibitem[{\citenamefont{Albert et~al.}(1999)\citenamefont{Albert, Jeong, and
  Barab\'{a}si}}]{www99}
\bibinfo{author}{\bibfnamefont{R.}~\bibnamefont{Albert}},
  \bibinfo{author}{\bibfnamefont{H.}~\bibnamefont{Jeong}}, \bibnamefont{and}
  \bibinfo{author}{\bibfnamefont{A.-L.} \bibnamefont{Barab\'{a}si}},
  \bibinfo{journal}{Nature} \textbf{\bibinfo{volume}{401}},
  \bibinfo{pages}{130} (\bibinfo{year}{1999}).

\bibitem[{\citenamefont{Albert and Barab\'{a}si}(2002)}]{barabasi01}
\bibinfo{author}{\bibfnamefont{R.}~\bibnamefont{Albert}} \bibnamefont{and}
  \bibinfo{author}{\bibfnamefont{A.-L.} \bibnamefont{Barab\'{a}si}},
  \bibinfo{journal}{Rev. Mod. Phys.} \textbf{\bibinfo{volume}{74}},
  \bibinfo{pages}{47} (\bibinfo{year}{2002}).

\bibitem[{\citenamefont{Liljeros et~al.}(2001)\citenamefont{Liljeros, Edling,
  Amaral, Stanley, and Aberg}}]{amaral01}
\bibinfo{author}{\bibfnamefont{F.}~\bibnamefont{Liljeros}},
  \bibinfo{author}{\bibfnamefont{C.~R.} \bibnamefont{Edling}},
  \bibinfo{author}{\bibfnamefont{L.~A.~N.} \bibnamefont{Amaral}},
  \bibinfo{author}{\bibfnamefont{H.~E.} \bibnamefont{Stanley}},
  \bibnamefont{and} \bibinfo{author}{\bibfnamefont{Y.}~\bibnamefont{Aberg}},
  \bibinfo{journal}{Nature} \textbf{\bibinfo{volume}{411}},
  \bibinfo{pages}{907} (\bibinfo{year}{2001}).

\bibitem[{\citenamefont{Hethcote and Yorke}(1984)}]{het84}
\bibinfo{author}{\bibfnamefont{H.~W.} \bibnamefont{Hethcote}} \bibnamefont{and}
  \bibinfo{author}{\bibfnamefont{J.~A.} \bibnamefont{Yorke}},
  \bibinfo{journal}{Lect. Notes Biomath.} \textbf{\bibinfo{volume}{56}},
  \bibinfo{pages}{1} (\bibinfo{year}{1984}).

\bibitem[{\citenamefont{Anderson and May}(1992)}]{anderson92}
\bibinfo{author}{\bibfnamefont{R.~M.} \bibnamefont{Anderson}} \bibnamefont{and}
  \bibinfo{author}{\bibfnamefont{R.~M.} \bibnamefont{May}},
  \emph{\bibinfo{title}{Infectious diseases in humans}}
  (\bibinfo{publisher}{Oxford University Press}, \bibinfo{address}{Oxford},
  \bibinfo{year}{1992}).

\bibitem[{\citenamefont{Pastor-Satorras and
  Vespignani}(2001{\natexlab{a}})}]{pv01avir}
\bibinfo{author}{\bibfnamefont{R.}~\bibnamefont{Pastor-Satorras}}
  \bibnamefont{and}
  \bibinfo{author}{\bibfnamefont{A.}~\bibnamefont{Vespignani}},
  \bibinfo{journal}{Phys. Rev. Lett.} \textbf{\bibinfo{volume}{86}},
  \bibinfo{pages}{3200} (\bibinfo{year}{2001}{\natexlab{a}}).
\bibinfo{author}{\bibfnamefont{R.}~\bibnamefont{Pastor-Satorras}}
  \bibnamefont{and}
  \bibinfo{author}{\bibfnamefont{A.}~\bibnamefont{Vespignani}},
  \bibinfo{journal}{Phys. Rev. E} \textbf{\bibinfo{volume}{63}},
  \bibinfo{pages}{066117} (\bibinfo{year}{2001}{\natexlab{b}}).




\bibitem[{\citenamefont{May and Lloyd}(2001)}]{lloydsir}
\bibinfo{author}{\bibfnamefont{R.~M.} \bibnamefont{May}} \bibnamefont{and}
  \bibinfo{author}{\bibfnamefont{A.~L.} \bibnamefont{Lloyd}},
  \bibinfo{journal}{Phys. Rev. E} \textbf{\bibinfo{volume}{64}},
  \bibinfo{pages}{066112} (\bibinfo{year}{2001}).

\bibitem[{\citenamefont{Moreno et~al.}(2001)\citenamefont{Moreno,
  Pastor-Satorras, and Vespignani}}]{moreno}
\bibinfo{author}{\bibfnamefont{Y.}~\bibnamefont{Moreno}},
  \bibinfo{author}{\bibfnamefont{R.}~\bibnamefont{Pastor-Satorras}},
  \bibnamefont{and}
  \bibinfo{author}{\bibfnamefont{A.}~\bibnamefont{Vespignani}},
  \emph{\bibinfo{title}{Epidemic outbreaks in complex heterogeneous networks}}
  (\bibinfo{year}{2001}), \bibinfo{note}{e-print cond-mat/0107267}.

\bibitem[{\citenamefont{Pastor-Satorras and Vespignani}(2002)}]{psvpro}
\bibinfo{author}{\bibfnamefont{R.}~\bibnamefont{Pastor-Satorras}}
  \bibnamefont{and}
  \bibinfo{author}{\bibfnamefont{A.}~\bibnamefont{Vespignani}},
  \bibinfo{journal}{Phys. Rev. E} \textbf{\bibinfo{volume}{65}},
  \bibinfo{pages}{036104} (\bibinfo{year}{2002}).

\bibitem[{\citenamefont{Dezs{\"o} and Barab{\'a}si}(2001)}]{aidsbar}
\bibinfo{author}{\bibfnamefont{Z.}~\bibnamefont{Dezs{\"o}}} \bibnamefont{and}
  \bibinfo{author}{\bibfnamefont{A.-L.} \bibnamefont{Barab{\'a}si}},
  \emph{\bibinfo{title}{Can we stop the \mbox{AIDS} epidemic?}}
  (\bibinfo{year}{2001}), \bibinfo{note}{e-print cond-mat/0107420}.

\bibitem[{\citenamefont{Marro and Dickman}(1999)}]{marro99}
\bibinfo{author}{\bibfnamefont{J.}~\bibnamefont{Marro}} \bibnamefont{and}
  \bibinfo{author}{\bibfnamefont{R.}~\bibnamefont{Dickman}},
  \emph{\bibinfo{title}{Nonequilibrium phase transitions in lattice models}}
  (\bibinfo{publisher}{Cambridge University Press},
  \bibinfo{address}{Cambridge}, \bibinfo{year}{1999}).

\bibitem[{\citenamefont{Dorogovtsev and Mendes}(2001)}]{dorogorev}
\bibinfo{author}{\bibfnamefont{S.~N.} \bibnamefont{Dorogovtsev}}
  \bibnamefont{and} \bibinfo{author}{\bibfnamefont{J.~F.~F.}
  \bibnamefont{Mendes}}, \emph{\bibinfo{title}{Evolution of networks}}
  (\bibinfo{year}{2001}), \bibinfo{note}{e-print cond-mat/0106144}.

\bibitem[{\citenamefont{Diekmann and Heesterbeek}(2000)}]{epidemics}
\bibinfo{author}{\bibfnamefont{O.}~\bibnamefont{Diekmann}} \bibnamefont{and}
  \bibinfo{author}{\bibfnamefont{J.}~\bibnamefont{Heesterbeek}},
  \emph{\bibinfo{title}{Mathematical epidemiology of infectious diseases: model
  building, analysis and interpretation}} (\bibinfo{publisher}{John Wiley \&
  Sons}, \bibinfo{address}{New York}, \bibinfo{year}{2000}).

\bibitem[{\citenamefont{Abramowitz and Stegun}(1972)}]{abramovitz}
\bibinfo{author}{\bibfnamefont{M.}~\bibnamefont{Abramowitz}} \bibnamefont{and}
  \bibinfo{author}{\bibfnamefont{I.~A.} \bibnamefont{Stegun}},
  \emph{\bibinfo{title}{Handbook of mathematical functions.}}
  (\bibinfo{publisher}{Dover}, \bibinfo{address}{New York},
  \bibinfo{year}{1972}).

\bibitem[{\citenamefont{Cohen et~al.}(2001)\citenamefont{Cohen, Erez,
  {ben-Avraham}, and Havlin}}]{havlin01}
\bibinfo{author}{\bibfnamefont{R.}~\bibnamefont{Cohen}},
  \bibinfo{author}{\bibfnamefont{K.}~\bibnamefont{Erez}},
  \bibinfo{author}{\bibfnamefont{D.}~\bibnamefont{{ben-Avraham}}},
  \bibnamefont{and} \bibinfo{author}{\bibfnamefont{S.}~\bibnamefont{Havlin}},
  \bibinfo{journal}{Phys. Rev. Lett.} \textbf{\bibinfo{volume}{86}},
  \bibinfo{pages}{3682} (\bibinfo{year}{2001}).

\bibitem[{\citenamefont{Callaway et~al.}(2000)\citenamefont{Callaway, Newman,
  Strogatz, and Watts}}]{newman00}
\bibinfo{author}{\bibfnamefont{D.~S.} \bibnamefont{Callaway}},
  \bibinfo{author}{\bibfnamefont{M.~E.~J.} \bibnamefont{Newman}},
  \bibinfo{author}{\bibfnamefont{S.~H.} \bibnamefont{Strogatz}},
  \bibnamefont{and} \bibinfo{author}{\bibfnamefont{D.~J.} \bibnamefont{Watts}},
  \bibinfo{journal}{Phys. Rev. Lett.} \textbf{\bibinfo{volume}{85}},
  \bibinfo{pages}{5468} (\bibinfo{year}{2000}).

\end{thebibliography}
\end{document}